%% file: main.tex
\documentclass[letterpaper,10pt]{article}

\usepackage[T1]{fontenc}
\usepackage{fontspec}
\usepackage{csquotes}
\usepackage[english]{babel}
\usepackage[
  left=0.8in,
  right=0.8in,
  top=1in,
  bottom=1in]{geometry}
\usepackage{setspace}
\usepackage[style = chem-acs, backend=biber]{biblatex}
\usepackage{graphicx}
\usepackage{float}
\newfloat{scheme}{htbp}{los}
\floatname{scheme}{Scheme}
\floatname{chart}{Chart}
\newfloat{graph}{htbp}{loh}

\usepackage{chemformula} 
\usepackage[version = 4]{mhchem} 

\usepackage{graphicx}
\usepackage{booktabs}
\usepackage[labelfont=bf]{caption}
\usepackage{subcaption}
\usepackage{amsmath, amssymb, amsthm, mathtools}
\usepackage{braket}
\usepackage{upgreek}
\usepackage{textgreek}
\usepackage[colorlinks,citecolor=blue,linkcolor=blue,urlcolor=blue]{hyperref}
\usepackage{bookmark}
\newcommand{\dd}{\mathrm{d}}

\newcommand{\abs}[1]{\left|#1\right|}

\usepackage{authblk}

\author[1]{Tom Hoekstra}
\author[1]{Sander A. Mann}
\author[1]{Jorik {van de Groep}*}
\affil[1]{Van der Waals-Zeeman Institute, Institute of Physics, University of Amsterdam, 1098 XH, Amsterdam, the Netherlands}

\title{\textbf{Finite-Size Effects in Nonlocal Metasurfaces}}
\date{*Email: j.vandegroep@uva.nl}

\begin{document}
\twocolumn[
\maketitle

\begin{center}\bfseries Abstract\end{center}
\begin{quotation}\noindent
Metasurfaces leveraging nonlocal resonances enable narrowband spectral control and strong near-fields, with applications spanning augmented reality, biosensing, and nonlinear optics. However, the large spatial extent of these modes also poses new challenges: finite-size effects often deteriorate the performance of practical, footprint-limited devices. Here, we develop a spatiotemporal coupled-mode theory model that intuitively and quantitatively captures how finite size affects the scattering response of nonlocal metasurfaces. This reveals that, when the modal propagation length becomes constrained by the physical interaction length, the scattered field shows strong interference fringes and linewidth broadening. We derive an expression for the quality factor that incorporates an additional edge-loss channel, demonstrating that the stored energy and effective lifetime scale exponentially with the interaction length. We validate these predictions experimentally using position- and momentum-resolved spectroscopy on a 30-µm-wide metasurface. Overall, this work formalizes the impact of finite size on the scattering response of nonlocal photonic systems, and provides handles on how to minimize the impact of finite-size effects in metasurface design.
\end{quotation}
\vspace{1em}
\noindent\textbf{Keywords:} Nonlocal metasurfaces, guided-mode resonances, finite-size effects, spatiotemporal coupled-mode theory, 2D materials

\vspace{3em}
]

\begin{refsection}

\section{Introduction}
Over the past decades planar arrays of resonant nanostructures---metasurfaces---have evolved from simple optical components mimicking their bulk counterparts to complex systems performing a multitude of novel functions \cite{yu2014, kuznetsov2024}. Using a variety of materials and geometries to tailor the scattering response of dielectric or metallic nanostructures, local metasurfaces can sculpt the spatial phase profile of an impinging wavefront with deeply subwavelength resolution. Increasingly, though, optoelectronic technologies require narrowband selectivity and strong light–matter interactions that are difficult to achieve with deeply subwavelength meta-atoms, whose quality ($Q$) factors tend to be low. 

Nonlocal metasurfaces, which harness spatially extended modes such as guided-mode resonances \cite{bandiera2008,magnusson2016,kim2017,lawrence2020} and quasi-bound states in the continuum \cite{hsu2013,koshelev2018,jin2019}, have emerged as a promising alternative to local metasurfaces \cite{shastri2023}. These long-lived resonant states can support extremely high $Q$-factors \cite{chen2022, huang2023,fang2024}, thereby delivering tremendous spectral resolution and near-field enhancements. Recently, it was even shown that nonlocalities can be engineered with locally varying features to simultaneously command both the spectral and spatial degrees of freedom \cite{overvig2020, overvig2022, chai2023}. Combined, these properties enable wide-ranging applications in biomolecular sensing \cite{wang2021, kuhner2022, hu2023}, coherent light sources \cite{kodigala2017, ha2018, hwang2021}, free-space modulators \cite{benea2022,damgaard-carstensen2025, hoekstra2026}, mixed-reality eyewear \cite{song2021, malek2022}, nonlinear optics \cite{koshelev2019, liu2019,zograf2022}, optical image processing \cite{kwon2018, cordaro2019}, thermal metasurfaces \cite{overvig2021, deluca2025}, and wavefront shaping \cite{klopfer2022,lin2023, hail2023}. It is therefore compelling to embrace nonlocal metasurfaces in the move toward ultracompact optical devices.

In the pursuit of miniaturizing meta-optical devices, the lateral dimensions of metasurfaces are increasingly compressed to fit within limited footprints. Whereas the impact of finite sizes can largely be ignored for local metasurfaces, with only select works investigating them \cite{rodriguez2012,grepstad2013,yang2014,zundel2018}, the inherent delocalization of high-$Q$ resonances in nonlocal metasurfaces implies that finite-size effects can no longer be neglected \cite{taghizadeh2017, droulias2018, liu2019}. Indeed, not long after the introduction of guided-mode resonant gratings \cite{wang1990, wang1993}, it was realized that finite sizes can degrade their performance \cite{brazas1995, saarinen1995, boye2000, jacob2000, bendickson2001, jeong2002, peters2004}. While these early works established the importance of finite-size effects, they were often limited to empirical observations or system-specific approximations, lacking a unified modeling framework. This trend persists today, as nonlocal metasurfaces are still typically simulated as infinitely periodic because finite arrays come with excessive computational costs in full-wave solvers and are not naturally captured using temporal coupled-mode theory \cite{haus1984, fan2003}. As a result, the resonant response of fabricated devices is often influenced by the unpredictable corollaries of finite size, which has already motivated mitigation strategies \cite{kintaka2012,taghizadeh2017,dolia2024,hao2025} and theoretical investigations \cite{ustimenko2024,hoang2025}. Nevertheless, a broadly applicable predictive modeling framework has so far remained elusive.

\begin{figure}[t]
    \centering
    \includegraphics[width=\columnwidth]{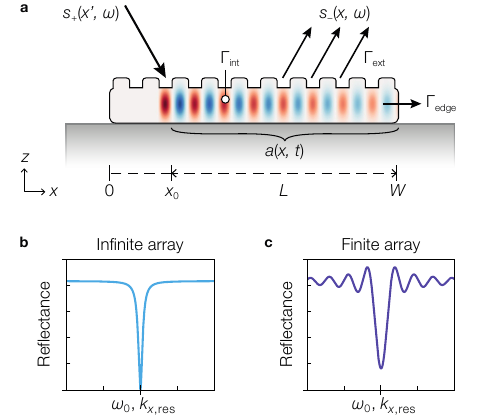}
    \caption{\textbf{A nonlocal metasurface with finite-size effects. (a)} An incident field $s_+(x',\omega)$ centered at $x_0$ excites a right-propagating quasi-guided mode with amplitude $a(x,t)$. The mode has an interaction length $L$, over which it decays through intrinsic loss ($\Gamma_\mathrm{int}$) and re-radiation ($\Gamma_\mathrm{ext}$). Power reaching the right edge, $x=W$, is irreversibly lost, giving rise to an additional edge loss channel $\Gamma_\mathrm{edge}$. The scattered field is denoted $s_-(x,\omega)$.
    \textbf{(b)} Modeled reflectance for an infinite, and
    \textbf{(c)} a finite metasurface array, respectively, showing the impact of finite size on the scattering response near the resonant frequency $\omega_0$ or in-plane wavevector $k_{x,\mathrm{res}}$.  
    }
    \label{fig1}
\end{figure}

Here, we develop a model that intuitively and quantitatively captures how a finite lateral footprint reshapes the scattering response of guided-mode resonant metasurfaces. We use spatiotemporal coupled-mode theory (STCMT), a recent extension of temporal coupled-mode theory that captures spatial inhomogeneity \cite{bykov2015, overvig2024, jeon2025}. Our model explicitly accounts for excitation of a traveling wave that can re-radiate within a finite aperture due to the limited metasurface width. We show that when the physical interaction length becomes comparable to the modal propagation length, finite-size effects fundamentally modify the optical response, giving rise to excitation position-dependent interference fringes and linewidth broadening. Crucially, we derive an expression for the $Q$-factor that incorporates an additional dissipation channel due to edge losses, demonstrating that the stored energy and effective lifetime scale exponentially with the interaction length $L$, defined as the distance over which the nonlocal mode travels after excitation at $x_0$ before reaching the termination (Fig.~\ref{fig1}a). We validate the model experimentally by fabricating a 30 µm wide metasurface and performing position- and momentum-resolved reflectance spectroscopy. The measurements show remarkable agreement with the theory, directly confirming the predicted size effects and enabling quantitative separation of the intrinsic, radiative, and edge-induced loss rates. Altogether, this work establishes STCMT as an invaluable tool for designing and interpreting footprint-limited nonlocal metasurfaces.

\section{Results}

\begin{figure*}[!t]
    \centering
    \includegraphics[width=\textwidth]{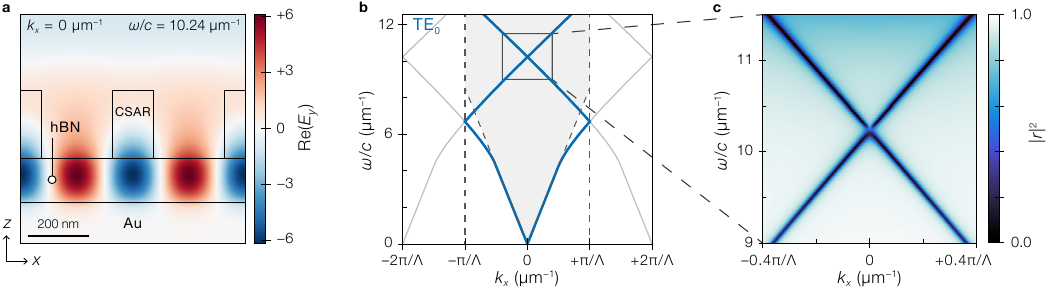}
    \caption{\textbf{Semi-infinite guided-mode resonator. (a)} Full-wave simulation of the electric-field profile $\mathrm{Re}(E_y)$ of the quasi-guided mode in a semi-infinite resonator comprised of a CSAR subwavelength grating on a hBN waveguide with an Au back-reflector.
    \textbf{(b)} Calculated TE\textsubscript{0} guided-mode dispersion, where the grating is modeled as an effective medium. The grating folds the dispersion into the first Brillouin zone ($-\pi/\Lambda$ to $+\pi/\Lambda$), shown by vertical dashed lines, with the portion that lies within the light cone indicated by the gray shaded region.
    \textbf{(c)} Simulated momentum-resolved reflectance ($|r|^2$) showing the corresponding dispersion.}
    \label{fig2}
\end{figure*}

In this work, we study the effects of finite size on the optical response of a reflective guided-mode resonator. Here, the structure consists of a truncated waveguide of width $W$ patterned with a subwavelength grating of period $\Lambda$ and backed by a reflector (Fig.~\ref{fig1}a). The scattering problem is described in terms of a single modal parameter $a(x,t)$ that is excited from free-space by a spatially localized input field $s_+(x')$. Compared to an infinite metasurface, lattice truncation can fundamentally alter the scattered field $s_-(x)$ near the resonant frequency $\omega_0$ or, equivalently, the in-plane wavevector $k_{x,\mathrm{res}}$, introducing fringes and broadening the resonance (Fig.~\ref{fig1}b,c).

To determine the origin of the finite size effects observed in Fig.~\ref{fig1}c, we first start with an infinitely periodic system. We consider a quasi-guided mode that is characterized by a dispersion that is linearized around a point ($\omega_0$, $\beta_0$) as
\begin{equation}
    \label{eq:lin-disp}
    \Omega(\beta) \simeq \omega_0 - i \Gamma + v_g (\beta-\beta_0)
\end{equation} 
where $\beta$ is the phase constant, $v_g$ is the group velocity, and $\Gamma = \Gamma_\mathrm{int} + \Gamma_\mathrm{ext}$ is the rate at which the mode decays due to internal (non-radiative) and external (radiative) mechanisms, respectively (Fig.~\ref{fig1}a). To achieve such dispersion, we design a dielectric hexagonal boron nitride (hBN) waveguide on a gold (Au) back-reflector that supports the fundamental TE$_0$ mode (Fig.~\ref{fig2}a). The waveguide is capped with a low-index dielectric polymer, specifically chemically semi-amplified resist (CSAR), in which a subwavelength grating is patterned with period $\Lambda = 368$~nm and duty cycle of 36\%. In this system, the duty cycle governs the radiative coupling ($\Gamma_\mathrm{ext}$), while non-radiative losses ($\Gamma_\mathrm{int}$) are incurred evanescently in the Au layer. We optimize the geometry to achieve mirror-like background reflection over the relevant bandwidth.

The periodic perturbation in the CSAR layer breaks the in-plane translational symmetry and folds the dispersion of the guided mode into the light cone (Fig.~\ref{fig2}b). The subwavelength periodicity restricts scattering to the 0\textsuperscript{th}-order direct reflection channel, with the diffracted orders constituting both left- and right-propagating evanescent waves owing to the grating's symmetric design. From full-wave simulations (Fig.~\ref{fig2}c), we find that the coupling between these modes is negligible except in the vicinity of $k_{x}=0$~{\textmu m$^{-1}$}, and consequently we can treat them as independent linear modes over most of the relevant $k_x$ range (this is verified in Supplementary Fig.~\ref{figS_loss_comparison}). This is not a general result, and a parabolic dispersion arises at the band edge when the coupling between forward and backward propagating modes is non-negligible \cite{bykov2015}. We attribute the minimal mode coupling to their predominant confinement in the continuous waveguide layer (Fig.~\ref{fig2}a) and to the relatively weak scattering efficiency of the low-contrast grating.

The dispersion is obtained numerically using an eigenmode solver by modeling the grating as an effective medium, which directly yields the most of the physical parameters in the dispersion relation, Eq.~\eqref{eq:lin-disp}. However, since the eigenmode is computed for a closed system, we must obtain $\Gamma_\mathrm{ext}$ by fitting the simulated response of the driven structure (Fig.~\ref{fig2}c). While $\Gamma_\mathrm{ext}$ may in principle depend on frequency or be modified by mode hybridization, we find that a constant value of $\Gamma_\mathrm{ext} = 7.4\times10^{12}$~s$^{-1}$ is sufficient to accurately reproduce the full-wave response over the bandwidth considered here (Supplementary Fig.~\ref{figS_loss_comparison}). 

The decay rate sets the resonance lifetime and is closely related to the quality factor, which is formally defined as the ratio of stored energy $U$ in the resonator to the rate of energy dissipation $P_\mathrm{loss}$, 
\begin{equation}
    \label{eq:Q_inf}
    Q 
    \equiv \omega_0 \frac{U}{P_\mathrm{loss}}
    = \frac{\omega_0}{2\Gamma}
\end{equation}
as well as the modal propagation length,
\begin{equation}
    L_p = \frac{v_g}{\Gamma}
\end{equation}
which governs the characteristic length scale over which nonlocality must be considered (\emph{i.e.} the nonlocality length \cite{overvig2024}). When an impinging wave launches a right-propagating mode at $x'=x_0$, the resulting excitation interacts with the structure over a finite length $L=W-x_0$. In the semi-infinite limit, $L/L_p \rightarrow \infty$, the array size negligibly perturbs the resonant response (Fig.~\ref{fig1}b). Conversely, in the short-grating limit, $L \lesssim L_p$, finite-size effects critically impair the theoretical performance (Fig.~\ref{fig1}c). At $\omega/c = 9$~\textmu m$^{-1}$, we find $L_p = 11.9$~\textmu m and $Q=132$, which reduces to $L_p = 6.4$~\textmu m and $Q=96$ at $\omega/c = 11.5$~\textmu m$^{-1}$ due to increased optical losses in the Au layer at higher frequencies \cite{yakubovsky2017}. 

The infinitely periodic limit discussed so far is typically the starting point of metasurface design, but it is clear that if the lateral dimensions of the metasurface are close to the propagation length $L_p$, we may expect significant finite-size effects. In the following, we will discuss how to take those finite-size effects into consideration using STCMT.

\subsection*{Finite-size effects}
Generally, the spatiotemporal equation of motion for a system with one input port and a locally linear dispersion are written as \cite{overvig2024}
\begin{equation}
    \begin{split}
     \frac{\partial a(x,t)}{\partial t}
     + v_g \frac{\partial a(x,t)}{\partial x}
     &+i\!\left(\omega_0 - i\Gamma - v_g\beta_0\right)a(x,t)  \\
     &  = \int \!\kappa(x,x')\, s_+(x',t)\ \dd x'
    \end{split}
    \label{eq:eom_general}
\end{equation}
This equation describes the propagation of a single modal amplitude $a(x,t)$ that is excited from free-space by an incident wave $s_+(x',t)$ with coupling coefficient $\kappa$. The parameter $a(x,t)$ is normalized such that its absolute value squared gives the instantaneous stored energy in the nonlocal mode per unit area, while $|s_+|^2$ captures the input power density.  As the mode propagates along the structure, it radiates to free space (described by $d$) to build up the scattered field $s_-(x,t)$ (also normalized such that $|s_-|^2$ is the outgoing power density) in interference with the non-resonant background, described by $Cs_+(x',t)$ (additional details in Supplementary Section \ref{S1:stcmt}): 

\begin{equation}
    s_-(x)
    =\int \! \left( C(x,x') \, s_+(x')
    + d(x,x') \,a(x') \right) \dd x'
    \label{eq:io_general}
\end{equation}

Although the mode has a nonlocal spatial distribution, we assume scattering is purely localized \cite{overvig2024}. As a result, the scattering coefficients become proportional to $\delta(x-x')$ (Supplementary Section \ref{S1:stcmt}). In the steady state, Eqs. \eqref{eq:eom_general} and \eqref{eq:io_general} then reduce to
\begin{equation}
\label{eq:ode_maintext}
v_g\left(
\frac{\partial}{\partial x}
- i\tilde{\beta}
\right)a(x) = \kappa(x)s_+(x)
\end{equation}
\begin{equation}
s_-(x) = -s_+(x) + d(x)a(x)
\end{equation}
where we introduced the complex propagation constant $\tilde{\beta}=\beta + i \alpha$ with attenuation rate $\alpha = \Gamma / v_g$ in the standard waveguide form. In addition, we assumed a perfectly reflective off-resonant response over the bandwidth of interest, \emph{i.e.} $C=-1$. 

To apply the STCMT framework to finite-size effects that arise upon truncating the metasurface to a width of $W$, we simply take the coupling coefficients to be non-zero in space only over the same width,
\begin{equation}
\kappa(x)= d(x)^* = \sqrt{2\Gamma_\mathrm{ext}}\,e^{iqx}\Theta(x)\Theta(W-x)
\end{equation}
where $\sqrt{2\Gamma_\mathrm{ext}}$ is the familiar radiative coupling term from temporal coupled-mode theory \cite{haus1984} and the phase factor ($e^{iqx}$, with $q=2\pi/\Lambda$) encodes in-plane momentum matching between the guided mode and free-space radiation. Crucially, the Heaviside functions ($\Theta$) restrict coupling to the finite interval $x\in[0,W]$. Note that while we have the standard TCMT identity $|\kappa|^2=|d|^2 = 2\Gamma_\text{ext}$, we also have the unusual relationship $\kappa = d^*$, which is a consequence of the in- and out-coupling coefficients carrying opposite grating momentum. 

To capture the essential physics of interest here, we find that it is sufficient to model \emph{only} a single, right-propagating mode,
\begin{equation}
    \label{eq:mode}
    a(x) = \frac{\sqrt{2\Gamma_\mathrm{ext}}}{v_g}
    \int_0^x
    e^{i\tilde{\beta}(x-x')}
    e^{iqx'}\,
    s_+(x')\,\dd x'.
\end{equation}
While this simplification strictly speaking violates reciprocity (due to the absence of a backward propagating mode), we find this approach justified in the regime where the interaction between the forward and backward modes is negligible. We note that, in principle, the theory can readily account for coupled, counter-propagating modes at the cost of increased complexity \cite{bykov2015}. 

The preceding equations provide us with a method to solve the reflected field from a finite nonlocal metasurface. In general, the easiest way to solve for the reflection involves numerically integrating Eq.~\ref{eq:mode} to find the mode amplitude, and then inserting this into Eq.~\ref{eq:io_general}. In certain cases, however, these equations can readily be solved analytically: for example, when the resonance linewidth is much narrower than the NA of excitation ($\alpha \ll k_{x,\mathrm{max}}$), we can approximate the diffraction-limited input field as a $\delta$-function (the validity of this approximation is confirmed in Supplementary Fig.~\ref{figS2}). This collapses the reflection coefficient to a compact closed form,
\begin{equation}
    \label{eq:r(k_x)_main}
    r(k_x) =-1+\frac{2\Gamma_\mathrm{ext}}{v_g \left(\alpha+i\Delta_k\right)}\,
    \left(1-e^{-(\alpha+i\Delta_k)L}\right).
\end{equation}
or similarly in terms of $\omega$ (full derivation in Supplementary Section \ref{S2}). Here, the first factor recovers the familiar Lorentzian dependence on detuning $\Delta_k=k_x-k_{x,\mathrm{res}}$, while the exponential term encodes the effects of finite size via the interaction length $L$. The prevalence of these effects is governed by the relative magnitudes of $L$ and $L_p$. In the short-grating limit, $L \lesssim L_p$, the finite aperture produces oscillatory spectral features and apparent resonance broadening, whereas in the long-grating limit, $L\gg L_p$, the finite-size term vanishes and Eq.~\eqref{eq:r(k_x)_main} reduces to the standard one-port temporal coupled-mode expression \cite{haus1984}. 

\begin{figure*}[!t]
    \centering
    \includegraphics[width=\textwidth]{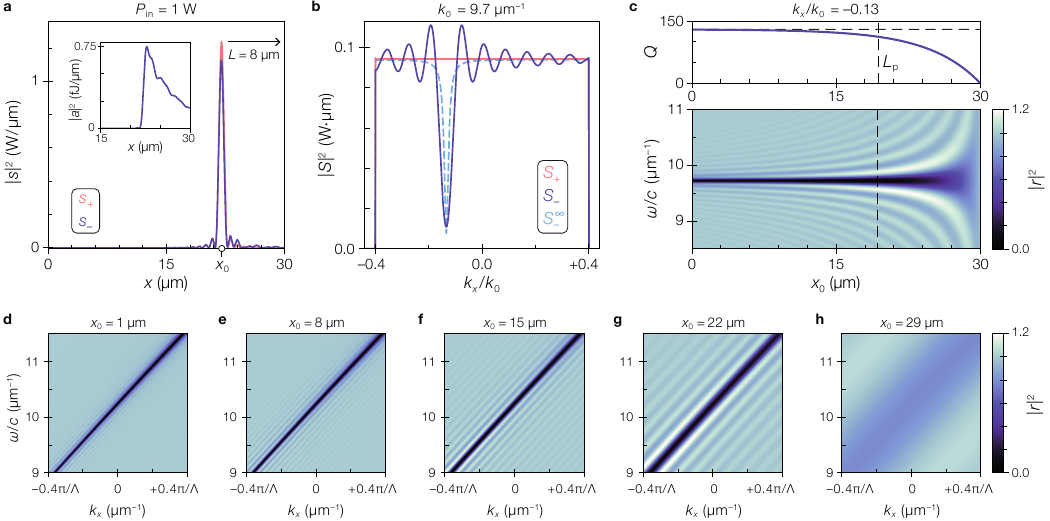}
    \caption{\textbf{Spatiotemporal coupled-mode theory of a finite-size guided-mode resonator. (a)} Real-space input field $|s_+(x')|^2$ (salmon) and the corresponding scattered field $|s_-(x)|^2$ (purple), calculated for a $P_\mathrm{in} = 1$~W excitation at $x_0=22~\mu\mathrm{m}$ ($L=8~\mu\mathrm{m}$) and $\mathrm{NA}=k_{x,\mathrm{max}}/k_0=0.4$. The inset shows the abrupt truncation of the mode amplitude $a(x)$ at $x=W$. \textbf{(b)} The corresponding transverse-momentum representation of the input field $|S_+(k_x)|^2$. The resulting scattered field $|S_-(k_x)|^2$ shows broadening and fringes arising from edge truncation when compared to the semi-infinite response $|S_{-}^{\infty}|^2$ (blue, dashed).
    \textbf{(c)} Evolution of the $Q$-factor (top) and reflectance (bottom) as a function of $x_0$, for fixed $k_x/k_0=-0.13$, with the propagation length $L_p$ indicated with respect to the edge of the resonator (vertical, dashed). The $Q$-factor tends to the semi-infinite limit (horizontal, dashed) as $L$ increases.
    \textbf{(d--h)} The corresponding reflectance dispersion showing $x_0$-dependent interference fringes and linewidths.
    }
    \label{fig3}
\end{figure*}

In the following, we apply the model to a finite metasurface with width $W=30$~\textmu m illuminated by a coherent, diffraction-limited input field arising from the Fourier transform of a one-dimensional entrance pupil with numerical aperture $\mathrm{NA}=k_{x,\max}/k_0=0.4$. The excitation launches a right-propagating quasi-guided mode that decays through absorption and radiation. As the mode reaches the termination of the metasurface at $x=W$, we assume that any remaining power is irreversibly lost. The finite width $W$ thus restricts the mode's interaction length, or scattering aperture, to $L=W-x_0$. Figure~\ref{fig3}a shows the input and output fields at frequency $k_0=\omega/c=9.7$~\textmu m$^{-1}$ and centered at $x_0 = 22$~\textmu m, corresponding to an interaction length $L=8$~\textmu m. We observe strongly localized excitation with characteristic sinc fringes (in salmon), and a reflected field that is only slightly reduced in amplitude (in purple), since most of the incident power is distributed among non-resonant $k$-vectors. However, when looking at the mode amplitude, we clearly observe an excitation that is launched at $x_0$, right-propagates and exponentially decays (Fig.~\ref{fig3}a, inset). At this frequency, the propagation length is $L_p = 10.6$~\textmu m such that the mode amplitude only reduces to $e^{-L/L_p}\approx 0.47$ before reaching the termination at $x=W$, and finite-size effects cannot be ignored. 

While, in real space, the effects of lattice truncation appear to be mild, the same cannot be said about Fourier space (Fig.~\ref{fig3}b). The angular scattering response still exhibits a resonant dip at $k_x/k_0=-0.13$ (in purple), consistent with the semi-infinite structure (in dashed blue), but the overall response deviates strongly from a simple Lorentzian lineshape. Instead, the resonant feature is modulated by a sinc-like envelope set by the finite scattering aperture, with pronounced fringes arising from coherent interference between radiation emitted from different positions along the grating. We note that $|s_-(k_x)|^2$ can locally exceed unity due to a redistribution of power within the angular spectrum and does not imply optical gain; the total scattered power is limited by the incident power minus the power dissipated through absorption and edge losses.

Moving the point of excitation along the sample changes $L$ and systematically reshapes the spectral (Fig.~\ref{fig3}h) and angular response (Fig.~\ref{fig3}d--g). As the beam approaches the right edge ($x_0\rightarrow W$), the interference pattern becomes more pronounced and the fringe spacing increases as $\Delta k_\mathrm{fringe} = 2\pi/L$, in direct analogy with Fraunhofer diffraction from a rectangular aperture. Conversely, moving $x_0$ farther away from the right edge reduces the fringe spacing. Ultimately, at the left edge ($L=29$ \textmu m), the response appears indistinguishable from the infinite array (Fig.~\ref{fig3}d). In this case, the normalized propagation length is $L/L_p = 2.7$ and the mode decays to $e^{(-L/L_p)}\approx 0.06$ before reaching $x=W$---providing an indication of the relative sizes required to obtain unperturbed performance.

Due to finite-size broadening, the $Q$-factor cannot be inferred directly from the spectral width of the resonant feature. A physically meaningful $Q$ should reflect the stored energy and energy losses of the guided mode. To capture the influence of edge losses, we introduce an effective decay rate $\Gamma_\mathrm{tot}(L)=\Gamma+\Gamma_\mathrm{edge}(L)$ where $\Gamma_\mathrm{edge}$ is the $L$-dependent edge loss rate. We can then rewrite Eq.~\eqref{eq:Q_inf} as
\begin{equation}
    \label{eq:Q_L}
    Q(L)
    = \frac{\omega_0}{2\Gamma_\mathrm{tot}(L)}
    = \frac{\omega_0}{2\Gamma}\left(1-e^{-2L/L_p}\right)
\end{equation}
which yields a lifetime-based $Q$ that formalizes the dependence on the interaction length and the propagation length, vanishing as $L\rightarrow0$ while approaching the semi-infinite limit ($Q\rightarrow130$ at $k_0=9.7$~\textmu m$^{-1}$) as $L \gg L_p$ (Fig.~\ref{fig3}b). Together, these findings underscore the detrimental impact that finite size can have, even at a moderate $Q$-factor. 


\subsection{Experimental validation}

\begin{figure*}[!t]
    \centering
    \includegraphics[width=\textwidth]{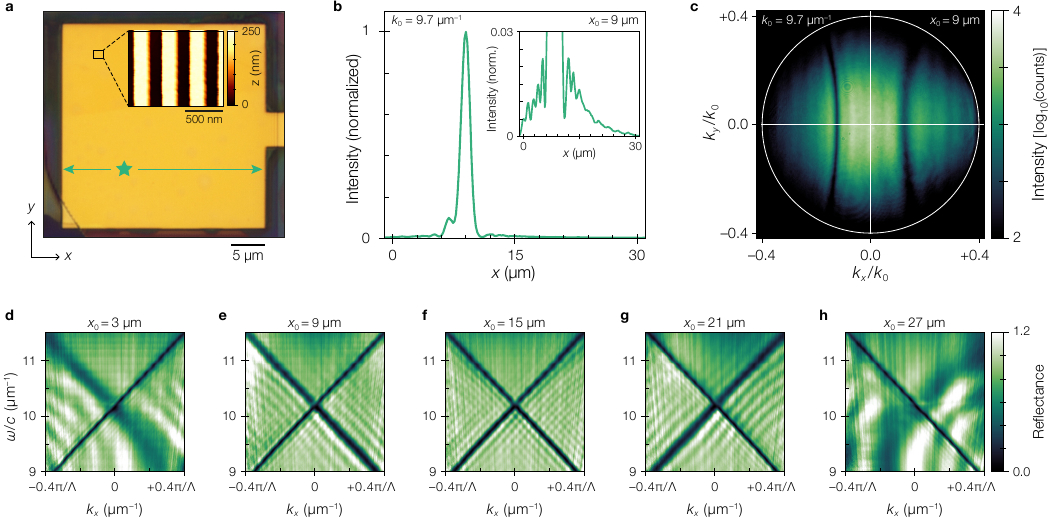}
    \caption{\textbf{Position-dependent linewidth and interference pattern. (a)} Optical micrograph of the fabricated resonator. Inset: atomic force micrograph of the subwavelength grating. The green star marks the excitation position ($x_0=9~\mu\mathrm{m}$). 
    \textbf{(b)} Horizontal linecut through the center of the reflected beam imaged in real-space. Inset: same image recorded with a longer exposure to reveal the low-intensity features, scaled to match the intensity in the main panel.
    \textbf{(c)} The corresponding reflection imaged in momentum space (Fourier plane). 
    \textbf{(d--h)} Measured momentum-resolved reflectance at different incident positions $x_0$, confirming the predicted interference fringes and linewidth variation.
    }
    \label{fig4}
\end{figure*}

To fabricate the designed metasurface, we use a dry-transfer technique \cite{castellanos2014} to stamp 143~nm-thick mechanically exfoliated hBN onto a prepatterned 30~\textmu m-wide Au back-reflector. We then spin-coat a 222~nm layer of CSAR and pattern the subwavelength grating via electron-beam lithography. Here, we utilize the etch-free approach---where the grating is patterned in a low-index resist rather than shallow-etched in the waveguide itself---to minimize fabrication imperfections and allow rapid prototyping \cite{huang2023, fang2024, shen2025, hoekstra2026}. Atomic force microscopy reveals excellent uniformity and confirms that the fabricated grating matches the design (Fig.~\ref{fig4}a). 

We mount the completed sample in a diffraction-limited optical microscope and illuminate it with a focused monochromatic laser beam (20$\times$ objective, $\mathrm{NA}=0.4$) at $k_0=9.7$~\textmu m$^{-1}$, centered at $x_0 = 9$~\textmu m. In contrast to the STCMT model, the symmetry of the grating leads to the excitation of two counter-propagating guided waves from the illumination spot. Figure~\ref{fig4}b shows a horizontal linecut through the real-space reflected intensity profile imaged on the camera. When recorded with a longer exposure time, a delocalized radiation pattern and exponential decay away from $x_0$ become apparent, characteristic of a leaky guided resonance (inset of Fig.~\ref{fig4}b).

By inserting a Bertrand lens, we image the back-focal (Fourier) plane of the microscope objective to obtain the scattered field in momentum space (Fig.~\ref{fig4}c). Two resonant features are observed as dips in the reflectance at $k_x/k_0\approx\pm0.13$. Because the excitation is off-center, the left- and right-propagating guided waves experience different interaction lengths. From the phase-matching condition of the forward mode, $k_x+q=\beta$, we obtain $k_x/k_0=-0.13$ and we therefore identify the feature at negative $k_x$ as right-propagating. This assessment is directly corroborated by the back-focal plane image, where this branch exhibits a narrower resonance than the branch at $k_x/k_0=+0.13$, owing to its longer interaction length $L$.

\begin{figure}[!t]
    \centering
    \includegraphics[width=\columnwidth]{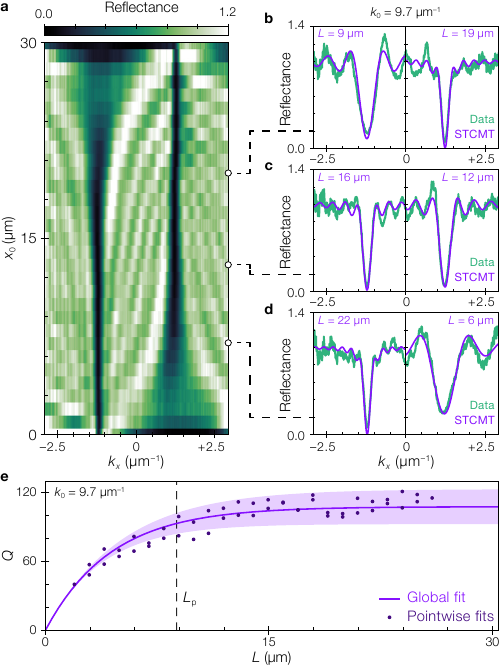}
    \caption{\textbf{Comparison between model and experiment. (a)} Measured reflectance versus excitation position $x_0$ and transverse momentum $k_x$, at $k_0=9.7~\mu\mathrm{m}^{-1}$. 
    \textbf{(b--d)} Representative momentum-resolved reflectance linecuts (green) at selected $x_0$, overlaid with the global STCMT fit (purple). The fitted lengths $L$ are indicated. 
    \textbf{(e)} Quality factor $Q(L)$ inferred from the global STCMT fit, with the extracted propagation length $L_p$ indicated (dashed line). Points represent fits performed independently at each linecut; the shaded band indicates the corresponding model discrepancy (one standard deviation) relative to these pointwise fits. 
    }
    \label{fig5}
\end{figure}

To study the position dependence of the resonant lineshapes in more detail, we scan the laser spot over the metasurface from $x=0$~\textmu m to $x=W=30$~\textmu m in 1~\textmu m steps. At each position, we sweep the laser frequency and record the momentum-resolved reflectance. Figures~\ref{fig4}d--h show the measured dispersion at five representative incident positions. When the metasurface is illuminated near its center (Fig.~\ref{fig4}f), the response is symmetric and both counter-propagating modes exhibit a narrow resonance while already showing an interference pattern indicating that the finite sample size affects the response. As the spot is moved toward the left (Fig.~\ref{fig4}e), this symmetry is broken: the right-propagating branch narrows (due to a longer $L$) while the left-propagating branch broadens, eventually becoming barely discernible near the edge (Fig.~\ref{fig4}d). This trend reverses when moving to the right edge (Fig.~\ref{fig4}g,h), consistent with the swapped interaction lengths experienced by the two guided waves. To further illustrate this modal evolution, we plot the position-dependent angular reflectance at a fixed $k_0=9.7$~\textmu m$^{-1}$ (Fig.~\ref{fig5}a), emphasizing how both the resonance lineshape and the interference fringes vary with the excitation position. Overall, these observations are in excellent agreement with the STCMT predictions (Fig.~\ref{fig3}).

Motivated by this qualitative correspondence, we next extract the intrinsic decay rates, $\Gamma_\mathrm{int}$ and $\Gamma_\mathrm{ext}$ by fitting the measured reflectance. To reduce the parameter space, we fix $\beta$ and $v_g$ from the eigenmode dispersion (Fig. \ref{fig2}b) and capture experimental deviations by a small $\Delta k_x$ offset. In addition, we introduce an effective width correction, $\Delta W$, to account for a systematic reduction of $L$ due to edge apodization in the fabricated metasurface. Figures~\ref{fig5}b--d show representative linecuts at three excitation positions $x_0$, with the STCMT fitted separately to the left- and right-propagating branches (since the STCMT was developed for a single mode only). The model accurately reproduces the reflectance data across the scanned range for $\Delta k_{x}\approx0.08$~{\textmu m\textsuperscript{-1}} and $\Delta W\approx-2.1$~{\textmu m}, including many of the smaller fringes around the resonant dips. From these fits, we obtain $\Gamma_\mathrm{ext} \approx 7.1\times10^{12}$~s$^{-1}$ and $\Gamma_\mathrm{int} \approx 6.5\times10^{12}$~s$^{-1}$, corresponding to a propagation length $L_p \approx 8.8$~{\textmu m}. Via Eq. \eqref{eq:Q_L}, we then retrieve the experimental $Q(L)$ curve with $Q\rightarrow107$ as $L\gg L_p$ (Fig.~\ref{fig5}e). Although the extracted radiative rate agrees closely with the modeled value $\Gamma^\mathrm{sim}_\mathrm{ext} = 7.4\times10^{12}$~s$^{-1}$, the non-radiative rate exceeds the modeled $\Gamma^\mathrm{sim}_\mathrm{int} = 3.9\times10^{12}$~s$^{-1}$ by almost a factor of two, which we attribute to additional dissipation due to surface roughness, polymer residues, and spatial inhomogeneity. As a result, the experimental $L_p$ and $Q$ are smaller than the modeled values (Fig. \ref{fig3}b). To nevertheless assess the model discrepancy, we re-fit $Q$ independently at each $x_0$ (data points in Fig.~\ref{fig5}e) and use this to define an uncertainty range on $Q$ (shaded region). These pointwise fits reproduce the same overall trend, further supporting the validity of our model.

\section{Design principles}
Although this work focused on a specific guided-mode resonant metasurface with a moderate $Q$-factor to demonstrate the impact of finite size, we stress that our observations apply more broadly. In principle, any propagating nonlocal mode that can be described by a locally linear dispersion will feature a $Q$-factor approaching the semi-infinite limit as $1 - e^{-2L/L_p}$. Concretely, our design advice for reaching the long-grating regime, in which finite-size effects can mostly be neglected, is to fabricate a metasurface with a physical width of $5\times$ the modal propagation length. This ensures that a mode excited in the center of the structure ($L/L_p=2.5$), decays to about 8\% of its original amplitude and the lifetime-based $Q$-factor reaches 99.3\% of the theoretical value. Of course, it may be interesting to operate in the limit of a size-constrained grating. Remarkably, in this regime the quality factor is primarily governed by the mode's transit time to the edge (Supplementary Section \ref{S3}). This implies that, for strongly footprint-constrained devices, high-$Q$ requires increasing the interaction time within the fixed footprint by engineering flat bands, \emph{i.e.} slow light \cite{taghizadeh2017, barton2021}.

\begin{figure*}[!t]
    \centering
    \includegraphics[width=\textwidth]{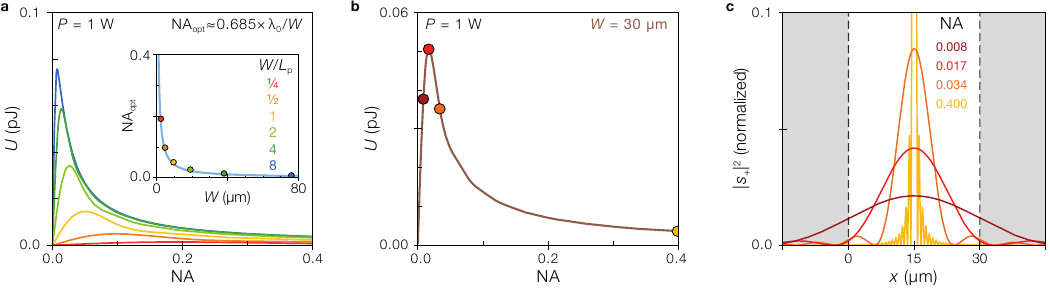}
    \caption{
    \textbf{Maximizing stored energy by matching illumination width to the grating. (a)} Stored energy ($U$) as a function of incident NA, calculated for varying grating widths $W/L_p$ (color) under sinc excitation in the center ($L=W/2$) with constant total incident power $P=1$~W. Inset: the optimal NA extracted at each width (colored datapoints) is well approximated by $\mathrm{NA}_\mathrm{opt} \approx 0.685\,\lambda_0/W$ (light blue curve).
    \textbf{(b)} Stored energy versus NA (brown curve) for the $W=30$~\textmu m grating discussed in the main text, with four highlighted datapoints corresponding to
    \textbf{(c)} the normalized incident intensity profiles $|s_+(x)|^2$ for the same four NA values. The grating boundaries are indicated (dashed lines); incident field outside the grating (gray shaded region) does not contribute to resonant energy build-up.
    }
    \label{fig6}
\end{figure*}

Another aspect that we have not examined so far is the impact of the NA (spot size) of the driving field, which relates to the experimental design rather than the sample size. In general, for an infinite grating, it is favorable to minimize the angular bandwidth such that all the incident power is concentrated at $k_{x,\mathrm{res}}$. This can be particularly important for applications requiring the highest possible light-matter interactions such as nonlinear optics. 

In the following, we consider a finite-NA sinc-beam to explore how the stored energy $U$ in the resonator is maximized by matching the spot size to the grating width (details in Supplementary Section \ref{S3}). Keeping the total incident power fixed at an arbitrary value of $P_\mathrm{inc} = 1~W$, we vary the NA to control the diffraction-limited spot size. We then evaluate 
\begin{equation}
    U\equiv\int_0^W\abs{a(x)}^2\ \mathrm{d} x
\end{equation}
by plugging the solution of the mode amplitude, Eq. \eqref{eq:mode} for various normalized widths $W/L_p$ (Fig.~\ref{fig6}a), observing a pronounced maximum that is captured well by the simple coherent aperture-overlap,
\begin{equation}
\int \! w(x)  s_+(x) \ \mathrm{d} x 
\end{equation}
Note that $U$ does not vary monotonically with NA, but rather features a maximum at some optimal NA value. For a sinc beam centered on the grating ($L=W/2$), we find the optimum to be $\mathrm{NA}\mathrm{opt} \approx 0.685 \lambda_0/W$. While the numerical prefactor depends on the specific excitation conditions assumed here, the overall scaling $\mathrm{NA}_\mathrm{opt}\propto \lambda_0/W$ holds generally.

To explore the origin of this maximum, let us again consider the 30-\textmu m-wide metasurface discussed throughout this work (Fig. \ref{fig6}b). We choose four representative numerical apertures to show that for a finite grating, as $\mathrm{NA}\rightarrow0$ the illumination profile $s_+(x)$ starts to extend well beyond the grating window $[0,W]$, and an increasing fraction of the incident power cannot excite the resonant mode. On the other hand, if the NA increases, an increasing fraction of the incident power is at wavenumbers outside of the resonant wavenumber, increasing reflection and reducing power coupled into the resonance. As a result, $U$ is maximized when the spot size is matched to the grating width, or vice versa, reflecting the trade-off between matching the real-space footprint and matching the resonance in momentum-space.

In sum, we have formalized the relations between the modal propagation length, physical metasurface footprint, excitation spot size, and critical performance metrics of nonlocal metasurfaces, which is especially relevant amid the recent push toward ultrahigh $Q$-factors. Our work provides practicable design guidelines that can be leveraged to reach the theoretical linewidth and maximize the light-matter interaction: i) to achieve a $Q$-factor close to that of the infinite array, the metasurface needs to be over five propagation lengths long, and ii) to maximize the stored power in the resonator, the array size must be matched to the central lobe of the incident beam. 

\section{Discussion and conclusion}
We present the effects of finite lateral extent on the optical properties of leaky-wave metasurfaces. In particular, we leverage the STCMT framework to derive approximate analytic expressions for the spectro-spatial scattering response of a truncated guided-mode resonator. This reveals that, when the interaction length is on the order of, or shorter than the guided wave's propagation length, the scattered field shows strong interference fringes that can exceed unity reflectance due to a redistribution of spectral weight in momentum space, as well as a reduction of the effective lifetime and therefore the energy stored in the resonator. To validate these results, we fabricate the described metasurface and perform position- and momentum-resolved reflectance spectroscopy. These measurements demonstrate the finite-size effects, observing excellent agreement between theory and experiment. More generally, our model offers a straightforward method to predict the maximum achievable quality factor in a limited footprint, given the theoretical dispersion of the guided wave. This further establishes STCMT as a valuable technique to both qualitatively and quantitatively capture the scattering response of real-world nonlocal meta-optical elements. 

The broad applicability of STCMT leaves many promising directions for further inquiry. For instance, future work could investigate the use of in-plane reflectors to increase the effective interaction length by trapping light within a finite scattering aperture \cite{kintaka2012,dolia2024}. Other directions include metasurfaces exhibiting a quadratic dispersion arising from quasi-bound states in the continuum or strongly coupled counter-propagating leaky waves, as well as two-port systems to model transmission-based meta-optics \cite{bykov2015, taghizadeh2017, cordaro2019, shastri2023}. Altogether, the framework presented here provides fundamental insight into the impact of finite size on the resonant response of these emerging photonic systems---an essential step toward technological integration in augmented reality, biosensing, and nonlinear optics.

\section*{Acknowledgements}
This work was funded by a Vidi grant (VI.Vidi.203.027) from the Dutch National Science Foundation (NWO). J.v.d.G. is also supported by a European Research Council Starting Grant under grant agreement No. 101116984. 

\section*{Data availability}
A full replication package including all data and scripts will be made openly available upon publication.

\section*{Conflict of interest}
The authors declare no competing interests.

\section*{Contributions}
T.H., J.v.d.G., and S.A.M. conceived the concept behind this research. S.A.M. and T.H. developed the STCMT model. T.H. fabricated the sample, and performed the measurements and simulations with input from J.v.d.G. and S.A.M. All authors contributed to analyzing the results and writing the manuscript.

\printbibliography[title={References}]
\end{refsection}

\clearpage
\onecolumn
\section*{Supporting Information}
\setcounter{figure}{0}
\renewcommand{\thefigure}{S\arabic{figure}}
\renewcommand{\thesection}{S\arabic{section}}
\setcounter{secnumdepth}{1}
\noindent\textbf{This file includes:}
\begin{itemize}
  \item Materials and methods
  \item Sections S1--4
  \item Figures S1--3
\end{itemize}
\clearpage
\begin{refsection}
\input{SI}

\printbibliography[title={References}]
\end{refsection}
\end{document}

%% file: SI.tex
\section*{Methods}
\subsubsection*{Sample fabrication} 
A bulk hBN crystal (HQ Graphene) is mechanically exfoliated using Nitto SPV-224 tape onto stamps cut from polydimethylsiloxane sheets (Gel-Pak WF-30). The stamps are inspected in a confocal microscope (WITec \textalpha-300) and hBN flakes of suitable thickness are identified using white-light reflectometry. The target substrate, with the prepatterned Au back-reflector, is mounted in a custom-built stamping microscope equipped with a temperature-controlled stage. After aligning the chosen hBN flake over the back-reflector region, the stamp is brought into contact with the substrate and held at 60~\textdegree C for 5~min. The stamp is subsequently retracted at 0.1~\textmu m/s, leaving the hBN flake on the substrate. To improve adhesion and reduce surface contamination, the sample is annealed in a vacuum tube furnace ($P \approx 10^{-6}$--$10^{-7}$~mbar) at 150~\textdegree C for 8~h. The surfaces is further cleaned with an O\textsubscript{2} plasma descum for 30~s (Oxford Instruments Plasma 80). A 222~nm layer of CSAR AR-P 62 e-beam resist is spin-coated and baked at 150~\textdegree C for 120~s. The grating is patterned by electron-beam lithography (Raith Voyager) using a dose of 130.8~\textmu C/cm\textsuperscript{2}. Development is performed in n-amyl acetate for 60~s, followed by sequential rinses in o-xylene (7~s) and MIBK:IPA 9:1 (15~s). The sample is finally rinsed in IPA and dried under N\textsubscript{2}. Atomic force microscopy (Bruker Dimension FastScan) is performed to verify the fabricated grating dimensions against the design.

\subsubsection*{Optical measurements}
Reflection measurements are performed with a diffraction-limited confocal microscope (WITec \textalpha-300). As the light source, we use an NKT Photonics SuperK tunable laser system coupled into the microscope via a single-mode photonic crystal fiber (Connect FD7-PM). Linearly polarized illumination is provided by a pulsed supercontinuum source (Extreme EXW-12; 78~MHz repetition rate, $\sim$5~ps pulse width), which is spectrally selected with an acousto-optic tunable filter (Select VIS/1X) to yield coherent monochromatic light with a bandwidth of $\sim$1~nm across the visible range. The collimated beam is directed to the entrance pupil (Fourier plane) of the objective (Zeiss EC Epiplan, $20\times$, $\mathrm{NA}=0.4$), producing a diffraction-limited spot at the sample plane. The sample is mounted on the microscope piezo stage for precise lateral positioning.

The reflected signal is collected by the same objective and imaged onto a 14-bit CMOS camera (Zeiss Axiocam 705 mono). A Bertrand lens can be inserted in the detection path to switch between real-space and momentum-space imaging. The incident power and camera exposure time are calibrated and adjusted slightly per wavelength to maximize the signal-to-noise ratio; the incident power is in the range 36--80~nW. For the position-dependent dispersion measurements, we automate wavelength sweeps and piezo-stage scans using a custom Python control script. Two reference measurements are acquired: one without a sample mounted to account for internal reflections, and one on an exposed gold region for normalization. The latter uses the theoretical angle- and wavelength-dependent reflectance of thin-film Au calculated from tabulated permittivity data \cite{yakubovsky2017}.

\subsubsection*{Numerical simulations}
The field distribution and reflectance dispersion of the infinite structure are simulated using S\textsuperscript{4} rigorous coupled-wave analysis \cite{victorliu2012}. The eigenmode dispersion is calculated with a custom slab waveguide mode solver, in which the grating is treated as an effective medium. For hBN and Au, we use tabulated optical constants \cite{yakubovsky2017, lee2019}, whereas for CSAR we extract the permittivity from spectroscopic ellipsometry measurements (J.A. Woollam VB-400). The spatiotemporal coupled-mode equations are derived analytically, with full details provided in Supplementary Sections \ref{S1:disp}, \ref{S1:stcmt} and \ref{S2}. All simulations are carried out using custom Python scripts.

\setcounter{equation}{0}

\clearpage
\section{Dispersion relation}
\label{S1:disp}
\begin{figure}[htbp]
    \centering
    \includegraphics[width=\linewidth]{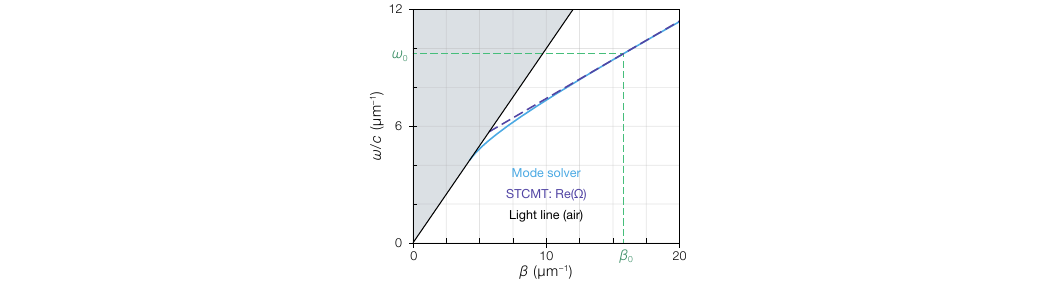}
    \caption{
    \textbf{Locally linear complex dispersion relation.} Simulated TE$_0$ eigenmode dispersion obtained with a slab waveguide mode solver by treating the grating as an effective medium (blue). Overlaid is the approximate STCMT dispersion $\Omega$ (purple, dashed) which is linearized around the point $\omega_0$, $\beta_0$ (green, dashed). For reference, the light cone (shaded gray) and light line in air (black) are shown.
    }
    \label{figS_disp}
\end{figure}

Consider a quasi-guided mode propagating in the positive $x$-direction described by the modal parameter $a(x,t)$, which evolves as
\begin{equation}
a(x,t)\propto e^{i\beta x -i\omega t}
\end{equation}
with phase constant $\beta$ and frequency $\omega$. Throughout, we approximate the dispersion as locally linear (Fig. \ref{figS_disp}), by expanding the complex propagation constant to first-order near a reference point ($\beta_0$,$\omega_0$) as  
\begin{equation}
    \label{eq:disp_beta}
\tilde{\beta}(\omega)=\beta+i\alpha \simeq \beta_0+\frac{\omega-\omega_0}{v_g}+i\alpha,
\qquad
v_g \equiv \left.\frac{d\omega}{d\beta}\right|_{\beta_0},
\end{equation} 
where $\alpha$ is the spatial attenuation rate. Equivalently, in the momentum-domain picture we can write the complex frequency 
\begin{equation}
    \label{eq:disp}
    \Omega(\beta)\equiv \omega-i\Gamma\simeq  \omega_0-i\Gamma + v_g(\beta-\beta_0),
\end{equation}
where $\Gamma \equiv \Gamma_\mathrm{ext}+\Gamma_\mathrm{int} = v_g \alpha$ is the temporal decay rate comprising the external (radiative) decay rate $\Gamma_\mathrm{ext}$ associated with coupling of the guided mode to the far field via the grating, and the intrinsic, non-radiative loss rate $\Gamma_\mathrm{int}$. A closely related parameter is the modal propagation length,
\begin{equation}
    L_p= \frac{1}{\alpha} = \frac{v_g}{\Gamma},
\end{equation}
\emph{i.e.}, the distance over which the mode amplitude decays to $1/e$. This parameter is particularly important because it governs the characteristic length scale over which finite-size effects cannot be ignored.

\begin{figure}[htbp]
    \centering
    \includegraphics[width=\linewidth]{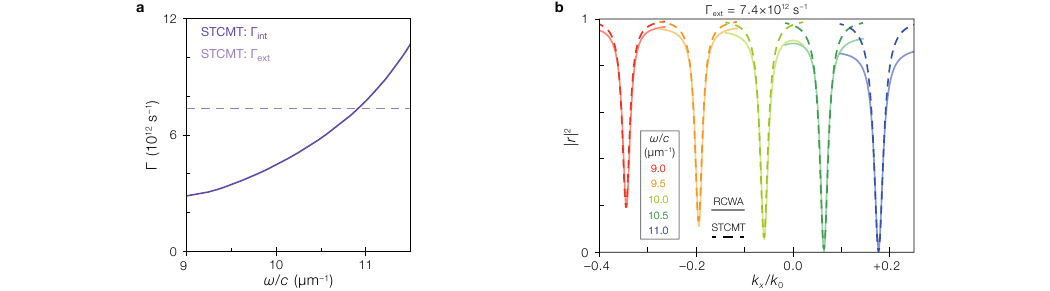}
    \caption{
    \textbf{Comparison with full-wave simulations.}
    \textbf{(a)} Loss rates used in STCMT: the eigenmode-solver decay rate is taken as the internal (non-radiative) loss rate $\Gamma_\mathrm{int}$ (purple, solid), while the external (radiative) loss rate $\Gamma_\mathrm{ext}$ is treated as approximately constant over the relevant frequency band and extracted by fitting to full-wave simulations (pink, dashed; $\Gamma_\mathrm{ext}=7.4\times10^{12}\,\mathrm{s^{-1}}$).
    \textbf{(b)} Momentum-resolved reflectance $|r(k_x)|^2$ from full-wave RCWA at selected frequencies (solid), overlaid with the corresponding STCMT prediction (dashed) evaluated in the long-grating limit. Colors indicate the frequencies $\omega/c$ as labeled.
    }
    \label{figS_loss_comparison}
\end{figure}

To parametrize the dispersion within STCMT, \emph{i.e.} Eq.~\eqref{eq:disp}, we obtain the guided mode dispersion numerically using a slab waveguide mode solver by modeling the grating as an effective medium. However, since the eigenmode is computed for a closed system, we must obtain the radiative rate $\Gamma_\mathrm{ext}$ by other means. In this case, we use full-wave simulations to model the driven response of the semi-infinite metasurface and fitting it with the well-known coupled-mode theory expression for a single mode with one external port (\emph{i.e.} Eq. \eqref{eq:r_inf}) \cite{haus1984, haus1991, fan2003}. From this analysis, we find that a constant value of $\Gamma_\mathrm{ext} = 7.4\times10^{12}$~s$^{-1}$ is sufficient to accurately reproduce the full-wave response over the bandwidth considered here (Fig.~\ref{figS_loss_comparison}). The observed deviation at higher frequencies is due to increasing optical losses in the gold layer, which is not accounted for in our model (constant background amplitude $C=-1$).

%
%

\section{Spatiotemporal coupled-mode equations}
\label{S1:stcmt}
\subsection*{Equation of motion}
The spatiotemporal dynamical equation for a system with a single mode $a(x,t)$ may be written as
\begin{equation}
    \label{eq:eom_general_SI}
     \frac{\partial a(x,t)}{\partial t}
     +i \Omega a(x, t)
      = \int \!\kappa(x,x')\, s_+(x',t)\ \mathrm{d} x',
\end{equation}
where $\kappa$ is the in-coupling parameter that couples the external driving field $s_+$ to the internal mode. Specifically, under the locally linear dispersion of Eq. \eqref{eq:disp}, this becomes
\begin{equation}
    \label{eq:eom_dispersion}
     \frac{\partial a(x,t)}{\partial t}
     +i\!\left(\omega_0 - i\Gamma + v_g(\beta-\beta_0)\right)a(x,t) 
      = \int \!\kappa(x,x')\, s_+(x',t)\ \mathrm{d} x'.
\end{equation}
Realizing the space-harmonic $\beta \rightarrow -i \partial_x$, this can be rewritten in the form presented in the main text:
\begin{equation}
     \frac{\partial a(x,t)}{\partial t}
     + v_g \frac{\partial a(x,t)}{\partial x}
     +i\!\left(\omega_0 - i\Gamma - v_g\beta_0\right)a(x,t) 
      = \int \!\kappa(x,x')\, s_+(x',t)\ \mathrm{d} x'.
\end{equation}
Instead realizing the time-harmonic $\partial_t \rightarrow - i \omega$, we can write the expression for the steady-state as a purely spatial differential equation:
\begin{equation}
     v_g \frac{\partial a(x)}{\partial x}
     +i\!\left(\omega_0 - \omega - i\Gamma - v_g\beta_0\right)a(x)
      = \int \!\kappa(x,x')\, s_+(x')\ \mathrm{d} x'.
\end{equation}
We can rewrite the left-hand side as
\begin{equation}
    \left(
    -i(\omega-\omega_0)
    + v_g \frac{\partial}{\partial x}
    + \Gamma
    - i v_g\beta_0
    \right)a(x)  =
    v_g\left(\frac{\partial}{\partial x} - i \tilde{\beta} \right) a(x) .
\end{equation}
Introducing the spatial operator $\mathcal{L}$,
\begin{equation}
    \mathcal{L} a(x) = v_g\left(\frac{\partial}{\partial x} - i \tilde{\beta} \right) a(x) ,
\end{equation}
we finally obtain the compact form
\begin{equation}
    \label{eq:ode:compact}
    \mathcal{L} a(x) = \int \!\kappa(x,x')\, s_+(x')\ \mathrm{d} x'.
\end{equation}

\subsection*{Mode amplitude}
To solve Eq. \eqref{eq:ode:compact}, we introduce the Green’s function $G(x;x')$, defined by, 
\begin{equation}
    \label{eq:greens_def}
    \mathcal{L} G(x;x') = \delta(x-x').
\end{equation}
For $x \neq x'$, the Green’s function satisfies
\begin{equation}
    \mathcal{L} G(x;x') = 0,
\end{equation}
which implies,
\begin{equation}
    \frac{\partial}{\partial x} G(x;x')
    =
    i\left(\beta_0 + \frac{\omega-\omega_0}{v_g} + i\frac{\Gamma}{v_g} \right)G(x;x')
    =
    i\tilde{\beta}\, G(x;x').
\end{equation}
The solution away from the source point therefore takes the form
\begin{equation}
    G(x;x') = A\,e^{i\tilde{\beta}(x-x')}, \qquad \text{for } x \neq x'.
\end{equation}
To determine the prefactor and the causal structure, we integrate Eq.~\eqref{eq:greens_def} over an infinitesimal interval around $x'$,
\begin{equation}
    \int_{x'-\epsilon}^{x'+\epsilon} \mathcal{L} G \, \mathrm{d} x
    =
    v_g\!\left[ G(x'+0;x') - G(x'-0;x') \right]
    = 1,
\end{equation}
where only the derivative term contributes to the discontinuity. Restricting to a right-propagating solution enforces causality,
$G(x<x')=0$, such that $G(x'-0;x')=0$. This yields $G(x'+0;x')=1/v_g$. The Green’s function can therefore be written as
\begin{equation}
    G(x;x') =
    \begin{cases}
        \dfrac{1}{v_g}\,e^{i\tilde{\beta}(x-x')} & \text{for } x>x',\\[6pt]
        0 & \text{for } x<x'.
    \end{cases}
\end{equation}
Equivalently, this can be expressed in compact form using the Heaviside step function $\Theta$,
\begin{equation}
    \label{eq:Greens}
    G(x;x') = \frac{1}{v_g}\,e^{i\tilde{\beta}(x-x')}\,\Theta(x-x').
\end{equation}
The mode amplitude follows from convolution of the Green’s function with the driving term,
\begin{equation}
    a(x)
    =
    \int_{-\infty}^{+\infty}
    \kappa(x')\,G(x;x')\,s_+(x')\,\mathrm{d} x',
\end{equation}
Finally, inserting Eq. \eqref{eq:Greens}, the mode amplitude is found as:
\begin{equation}
    \label{eq:mode_amplitude_solution}
    a(x)
    =
    \frac{1}{v_g}
    \int_{0}^x
    e^{i\tilde{\beta}(x-x')} \,
    \kappa(x')\,s_+(x')\mathrm{d} x'.
\end{equation}

\subsection*{Input-output relation}
The general form of the scattered field is written as
\begin{equation}
    s_-(x,t)
    =\int \! \left( C(x,x') \, s_+(x',t)
    + d(x,x') \,a(x',t) \right) \mathrm{d} x',
\end{equation}
where $C$ denotes the contribution from non-resonant (background) scattering and $d(x,x')$ is a out-coupling parameter which mediates coupling from the internal mode to external radiation. In this work, we assume spatially invariant background scattering and spatially instantaneous in- and out-coupling, such that 
\begin{equation}
    C(x, x') = C \delta(x-x'),
\end{equation}
\begin{equation}
    \kappa(x, x') = \kappa(x) \delta(x-x'),
\end{equation}
\begin{equation}
    d(x, x') = d(x) \delta(x-x').
\end{equation}
Under these assumptions, in the steady state, the input--output relation takes the form
\begin{equation}
    s_-(x) = C\,s_+(x) + d(x)\,a(x).
\end{equation}
The resonant contribution is generated by out-coupling of the guided mode back to the external channel. We write
\begin{equation}
\label{eq:sminus_res-realspace}
    s_-^{(\mathrm{res})}(x)=d(x)\,a(x).
\end{equation}
Coupling between the external field and the guided mode is enabled by a subwavelength grating of period $\Lambda$, which provides the grating momentum $q=2\pi/\Lambda$. In addition, coupling is restricted to a finite grating region of width $W$, which we represent by a window function
\begin{equation}
    w(x)=\Theta(x)\Theta(W-x),
\end{equation}
so that both in- and out-coupling vanish outside $x\in[0,W]$. Adopting this description, we model the in-coupling as
\begin{equation}
    \kappa(x)=\kappa_0\,w(x)\,e^{iqx},
\end{equation}
and the out-coupling as
\begin{equation}
    d(x)=d_0\,w(x)\,e^{-iqx}.
\end{equation}
The factors $e^{\pm iqx}$ enforce grating-assisted momentum matching: the external field acquires grating momentum upon coupling into the guided mode, and the opposite phase factor appears upon re-radiation to the external channel. The overall coupling strength is constrained by energy conservation \cite{fan2003, overvig2024}, such that $\kappa_0=d_0=\sqrt{2\Gamma_\mathrm{ext}}$, and
\begin{equation}
\label{eq:kappa}
\kappa(x)= d(x)^* = \sqrt{2\Gamma_\mathrm{ext}}\,e^{iqx}w(x).
\end{equation}
Finally, by plugging Eq. \eqref{eq:kappa} in Eq. \eqref{eq:mode_amplitude_solution}, we obtain the general expression for the mode amplitude for the metasurface under consideration:
\begin{equation}
    \label{eq:mode_SI}
    a(x) = \frac{\sqrt{2\Gamma_\mathrm{ext}}}{v_g}
    \int_0^x
    e^{i\tilde{\beta}(x-x')}
    e^{iqx'}\,
    w(x')\,
    s_+(x')\,\mathrm{d} x'.
\end{equation}


\section{Closed-form reflection under point excitation}
\label{S2}
\subsection*{Input field} 
In general, the incident field is modeled as a diffraction-limited spot centered at position $x_0$, which is represented in momentum-space as
\begin{equation}
    \label{eq:S+}
    S_+(k_x)=S_0\,e^{-ik_x x_0}, \quad \text{for } |k_x/k_0|\le \mathrm{NA},
\end{equation}
where $\lvert S_0\rvert^2$ represents the incident power per unit $k_x$. Here, the in-plane wavevector is $k_x=k_0 \sin \theta$ with free-space wavevector $k_0=\omega/c$ and incidence angle $\theta$. The numerical aperture $\mathrm{NA}=k_{x,\max}/k_0$ sets the excitation bandwidth.

In real space, this aperture-limited momentum-space distribution produces a sinc illumination profile,
\begin{equation}
    s_+(x')\propto \mathrm{sinc}\big(k_{x,\max}(x'-x_0)\big).
\end{equation}
This is analogous to a diffraction-limited optical system, where an Airy disk is produced by uniform illumination of the circular entrance pupil of a microscope objective. Crucially, when the width of the resonance in $k$-space (governed by $\alpha$) is sufficiently narrow compared to the numerical-aperture window ($\alpha / k_0 \ll \mathrm{NA}$), the incident field can be approximated as a point excitation localized at $x_0$,
\begin{equation}
    s_+(x') \approx S_0\,\delta(x'-x_0), \quad \text{with } x_0\in[0,W].
\end{equation}
This approximation enables the derivation of closed-form expressions for many relevant physical quantities within our framework. In the next section, we assess the accuracy of this approximation. We show that, for the system under consideration in the main text, the $\delta$-approximation closely reproduces the numerically calculated response under sinc illumination. 


\subsection*{Resonant scattered field}
In the following, we first derive a closed-form expression for the resonant contribution $S_-^{(\mathrm{res})}(k_x)$. We start from the formal solution of the mode amplitude, Eq.~\eqref{eq:mode_SI}, which, for the point-excitation approximation introduced above evaluates straightforwardly to
\begin{equation}
    a(x) = \frac{\sqrt{2\Gamma_\mathrm{ext}}\,S_0}{v_g}\,
    e^{i\tilde{\beta}(x-x_0)}
    e^{iqx_0}\,\Theta(x-x_0),
    \label{eq:mode-amplitude-closed}
\end{equation}
with the understanding that scattering is restricted to the grating region by $w(x)$. Plugging this into Eq. \eqref{eq:sminus_res-realspace}, we obtain
\begin{equation}
    s_-^{(\mathrm{res})}(x)=\frac{2\Gamma_\mathrm{ext}\,S_0}{v_g}\,
    e^{i(\tilde{\beta} - q)(x - x_0)}
    \Theta(x - x_0).
\end{equation}

To compute the resonant scattering contribution in $k$-space, we integrating from $x_0$ to the right edge $W$, \emph{i.e.}, over the modal interaction length $L=W-x_0$,
\begin{equation}
    S_-^{(\mathrm{res})}(k_x)=\int_{x_0}^{W} s_-^{(\mathrm{res})}(x)\,e^{-ik_xx}\,\mathrm{d} x.
\end{equation}
Substituting $u=x-x_0$ and rewriting yields
\begin{equation}
    S_-^{(\mathrm{res})}(k_x)=\frac{2\Gamma_\mathrm{ext}\,S_0}{v_g} e^{-ik_xx_0} \!
    \int_{0}^{L}
    e^{i\tilde{\beta}u}\,e^{-i(k_x + q)u}\,\mathrm{d} u.
\end{equation}
It is useful to define the detuning between the guided-mode wavevector and the input wave's in-plane momentum (for the $m=+1$ order) as
\begin{equation}
    \Delta_k=(k_\mathrm{x} + q) - \beta = k_x - k_{x,\mathrm{res}},
\end{equation}
where we additionally defined the resonant in-plane wavevector in free-space $k_{x,\mathrm{res}} = \beta-q$. Using this notation, we can write
\begin{equation}
    S_-^{(\mathrm{res})}(k_x)=\frac{2\Gamma_\mathrm{ext}\,S_0}{v_g}\,e^{-ik_xx_0}
    \int_{0}^{L}
    e^{-(\alpha+i\Delta_k)u}\mathrm{d} u,
\end{equation}
which evaluates to
\begin{equation}
    \label{eq:S-,res}
    S_-^{(\mathrm{res})}(k_x)=\frac{2\Gamma_\mathrm{ext}\,S_0}{v_g}\,e^{-ik_xx_0}
    \frac{1-e^{-(\alpha+i\Delta_k)L}}{\alpha+i\Delta_k}.
\end{equation}
\subsection*{Reflection coefficient}
The input-output relation in $k$-space is written as
\begin{equation}
    \label{eq:in-out-k}
    S_-(k_x) = CS_+(k_x) + S_-^{(\mathrm{res})}(k_x),
\end{equation}
and the reflection coefficient is defined as
\begin{equation}
    \label{eq:r(k_x)_def}
    r(k_x)\equiv \frac{S_-(k_x)}{S_+(k_x)}.
\end{equation}
For the metasurfaces considered here, the off-resonant response is mirror-like, such that $C\approx-1$ over the bandwidth of interest. The reflection coefficient is then readily obtained by plugging Eqs. \eqref{eq:S+} and \eqref{eq:S-,res} in Eq. \eqref{eq:r(k_x)_def}:
\begin{equation}
    \label{eq:r(k_x)}
    r(k_x) =-1+\frac{2\Gamma_\mathrm{ext}}{v_g \left(\alpha+i\Delta_k\right)}\,
    \left(1-e^{-(\alpha+i\Delta_k)L}\right).
\end{equation}
Under a linear dispersion,  $\Delta_\omega = v_g\Delta_k$, such that we can equivalently express the reflection coefficient in frequency-space for an ultrashort pulse as
\begin{equation}
    \label{eq:r(omega)}
    r(\omega) =-1+\frac{2\Gamma_\mathrm{ext}}{\Gamma+i\Delta_\omega}\,
    \left(1-e^{-(\Gamma+i\Delta_\omega)L/v_g}\right).
\end{equation}
In both representations, the response consists of the familiar Lorentzian dependence on the detuning as well as an exponential term which encodes the effects of a finite interaction length $L$. This length scale should be compared to the propagation length $L_p$ of the quasi-guided mode. When $L \lesssim L_p$, the finite aperture produces oscillatory spectral features and effective broadening of the resonance. On the other hand, in the long-grating limit $L\gg L_p$, the finite-size factor tends to unity and the reflection coefficient reduces to the well-known one-port TCMT expression \cite{haus1984},
\begin{equation}
    \label{eq:r_inf}
    r_\infty(\omega) = -1+\frac{2\Gamma_\mathrm{ext}}{\Gamma+i \Delta_\omega},
\end{equation}
and equivalently in $k$-space,
\begin{equation}
r_\infty(k_x) = -1+\frac{2\Gamma_\mathrm{ext}}{v_g(\alpha+i\Delta_k)}.
\end{equation}

\subsection*{Validity of the point excitation approximation}

\begin{figure}[ht]
    \centering
    \includegraphics[width=\linewidth]{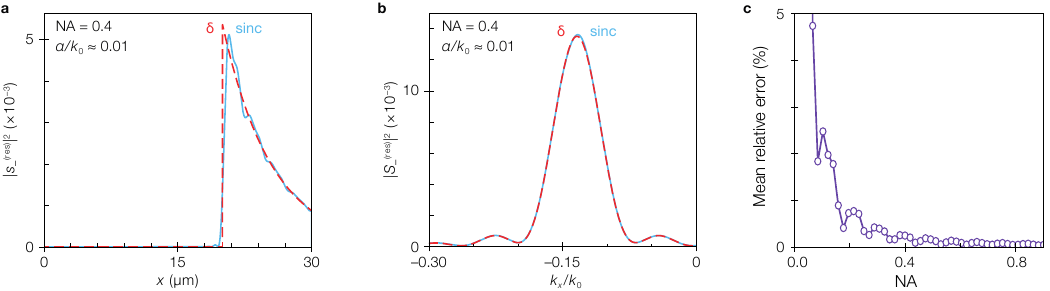}
    \caption{\textbf{Evaluating the $\delta$-function input approximation.} Comparison of the $\mathrm{NA}=0.4$ $\mathrm{sinc}$ excitation (blue, solid) and the closed-form expressions of the infinite-NA $\delta$-function excitation (red, dashed).
    \textbf{(a)} Resonant scattered field in real space, and
    \textbf{(b)} in momentum space evaluated at $k_0=9.7$~\textmu m$^{-1}$ and $L=10$~\textmu m, where $\alpha / k_0 \approx 0.01$ 
    \textbf{(c)} The mean relative error between $R(k_x)$ calculated using the sinc and $\delta$-function excitations, as function of NA, shows that the error diverges as $\mathrm{NA} \rightarrow \alpha/k_0$.
    }
    \label{figS2}
\end{figure}

\noindent To show that the $\delta$-approximation closely reproduces the numerically calculated response under sinc-illumination for the system under consideration, with $\mathrm{NA}=0.4$, we compare three key parameters with and without the $\delta$-approximation. Figures \ref{figS2}a,b show the resonant scattered field in real-space and in momentum-space, respectively, at $k_0=9.7$~\textmu m$^{-1}$, where the interaction length is $L=10$~\textmu m and the modal propagation length $L_p=10.6$~\textmu m, corresponding to $\alpha / k_0 \approx 0.01$ and thus satisfying $\alpha/k_0 \ll \mathrm{NA}$. While the real-space decay shows the biggest difference with a distinct oscillatory character arising from the sinc fringes, the momentum-space scattered field agrees almost perfectly. We further study the mean relative error between the reflectance ($|r(k_x)|^2$) calculated for the finite-NA sinc and infinite-NA $\delta$-function excitations (Fig. \ref{figS2}c). Notably, the error exhibits a sinc-like shape, remaining below 1\% when the NA is $\sim16\times$ larger than $\alpha/k_0$, while growing very large as $\mathrm{NA} \rightarrow \alpha/k_0$. Together, these results confirm the validity of the $\delta$-approximation for the system under consideration.

\section{Finite-size effects}
\label{S3}
\subsection*{Resonant anti-reflection}
A vanishing reflectance ($R=|r|^2=0$) corresponds to a destructive-interference condition in the output channel, which is modified by finite size. At resonance ($\Delta_k=0$), the finite interaction length $L$ introduces the factor $1-e^{-\alpha L}$ in the resonant scattering amplitude, and the well-known critical coupling criterion $\Gamma_\mathrm{ext}=\Gamma_\mathrm{int}$ no longer corresponds to $R=0$. Instead, the anti-reflection condition becomes
\begin{equation}
\label{eq:finiteL_R0_condition}
\frac{\Gamma_\mathrm{ext}}{\Gamma}
=\frac{1}{2\left(1-e^{-\alpha L}\right)},
\end{equation}
which generally requires stronger external coupling (more over-coupling) than the infinite-length case.

Solving Eq.~\eqref{eq:finiteL_R0_condition} for $L$ yields
\begin{equation}
\label{eq:finiteL_LR0}
L
= -\frac{v_g}{\Gamma}\,
\ln\!\left(1-\frac{\Gamma}{2\Gamma_\mathrm{ext}}\right).
\end{equation}
Given $\Gamma \geq 0$, a necessary condition for a real, positive solution of \eqref{eq:finiteL_LR0} is that $\Gamma_\mathrm{ext}>\Gamma/2$ (equivalently $\Gamma_\mathrm{ext}>\Gamma_\mathrm{int}$), otherwise no finite $L$ can produce $R=0$. It should be noted that, contrary to the infinite-length limit, the combination of parameters that yields perfect anti-reflection need not coincide with that which maximizes the stored energy $U$ in the resonator. 

\subsection*{Stored energy}
We next derive a closed-form expression for the stored energy
\begin{equation}
    U\equiv\int\abs{a(x)}^2\ \mathrm{d} x,
\end{equation}
for $x \in [0, W]$. Using the definition of $a(x)$ (Eq. \eqref{eq:mode-amplitude-closed}), we have
\begin{equation}
    \label{eq:mode-energy-density}
    \abs{a(x)}^2
    = 2\Gamma_\mathrm{ext} \left(\frac{S_0}{v_g}\right)^2 
    e^{-2\alpha(x-x_0)}\,\Theta(x-x_0).
\end{equation}
Since the excitation point lies within the grating ($x_0\in[0,W]$), the integral reduces to the interval $x\in[x_0,W]$:
\begin{equation}
    U = 2\Gamma_\mathrm{ext} \left(\frac{S_0}{v_g}\right)^2
    \int_{x_0}^{W} e^{-2\alpha(x-x_0)}\,\mathrm{d} x.
\end{equation}
Substituting $u=x-x_0$ and defining $L=W-x_0$ gives
\begin{equation}
    U = 2\Gamma_\mathrm{ext} \left(\frac{S_0}{v_g}\right)^2 \int_{0}^{L} e^{-2\alpha u}\,\mathrm{d} u
    = 2\Gamma_\mathrm{ext} \left(\frac{S_0}{v_g}\right)^2 \frac{1-e^{-2\alpha L}}{2\alpha}.
\end{equation}
Therefore, the stored energy is
\begin{equation}
    \label{eq:stored-energy}
    U = \frac{S_0^2}{v_g} \frac{\Gamma_\mathrm{ext}}{\Gamma} \left(1-e^{-2\alpha L}\right).
\end{equation}
In the long-grating limit $L\gg L_p$, the stored energy saturates to
\begin{equation}
    U \rightarrow \frac{S_0^2}{v_g} \frac{\Gamma_\mathrm{ext}}{\Gamma}.
\end{equation}

\subsection*{Edge loss}
The finite grating length introduces an additional loss channel associated with leakage of the quasi-guided wave at the grating edge.
We assume that the guided mode propagates only in the positive $x$-direction and that any power reaching the right edge $x=W$ is irreversibly lost, with no reflection back into the guided mode.

The guided power flux at position $x$ is proportional to
\begin{equation}
    P(x) \propto v_g \abs{a(x)}^2 .
\end{equation}
Using the expression for the mode amplitude derived above (Eq. \eqref{eq:mode-energy-density}), the power reaching the right edge is therefore
\begin{equation}
    P_\mathrm{edge} \propto v_g \abs{a(W)}^2
    = 2\Gamma_\mathrm{ext}\frac{S_0^2}{v_g} e^{-2\alpha L},
\end{equation}
where $L=W-x_0$ is the distance from the excitation point to the right edge.

It is convenient to express this edge loss in terms of an effective decay rate $\Gamma_\mathrm{edge}$, defined via the standard coupled-mode relation
\begin{equation}
    P_\mathrm{edge} = 2\Gamma_\mathrm{edge} U,
\end{equation}
with $U$ the stored energy obtained in the previous section (Eq. \eqref{eq:stored-energy}).
Substituting the closed-form expressions for $P_\mathrm{edge}$ and $U$ yields
\begin{equation}
    \label{eq:edge-loss}
    \Gamma_\mathrm{edge}
    = \Gamma \frac{e^{-2\alpha L}}{1-e^{-2\alpha L}}
    = \frac{\Gamma}{e^{2\alpha L}-1}.
\end{equation}

This expression highlights the exponential suppression of edge loss with increasing grating length.
In the long-grating limit ($L \gg L_p$), the edge contribution becomes negligible, whereas in the short-grating limit ($L \ll L_p$) the edge loss is dominated by the transit time to the boundary. Within this picture, edge loss constitutes an additional, geometry-induced decay channel that is distinct from intrinsic and radiative losses.

\subsection*{Quality factor}

While the finite grating length gives rise to sinc-like broadening in the $k$-space reflectance spectrum, this effect is geometric in origin and does not correspond to an intrinsic energy-decay process. A physically meaningful quality factor must therefore be defined in terms of the energy lifetime of the guided mode rather than the apparent spectral width of $|r(k_x)|^2$.

The quality factor of a resonance at angular frequency $\omega_0$ is defined as
\begin{equation}
    Q \equiv \omega_0 \frac{U}{P_\mathrm{loss}}
    = \frac{\omega_0}{2\Gamma_\mathrm{tot}},
\end{equation}
where $U$ is the stored energy, $P_\mathrm{loss}$ is the rate of energy dissipation, and $\Gamma_\mathrm{tot}$ is the total energy-decay rate.

For an infinitely long grating, the total decay rate is simply $\Gamma=\Gamma_\mathrm{ext}+\Gamma_\mathrm{int}$. For a finite grating of length $L$, however, we must include the additional edge loss channel. As shown above, this edge-induced loss can be expressed as an effective decay rate (Eq. \eqref{eq:edge-loss}), such that the total decay rate becomes
\begin{equation}
    \Gamma_\mathrm{tot}(L)
    = \Gamma + \Gamma_\mathrm{edge}(L)
    = \frac{\Gamma}{1-e^{-2\alpha L}}.
\end{equation}
The corresponding quality factor is therefore given by the closed-form
expression
\begin{equation}
    Q(L)
    = \frac{\omega_0}{2\Gamma_\mathrm{tot}(L)}
    = \frac{\omega_0}{2\Gamma}\left(1-e^{-2L/L_p}\right).
\end{equation}
This definition yields a lifetime-based quality factor that is not affected by apparent linewidth broadening arising from the finite scattering aperture.

In the long-grating limit ($L \gg L_p$), the edge contribution vanishes and the quality factor approaches its intrinsic value
\begin{equation}
    Q \rightarrow \frac{\omega_0}{2\Gamma}.
\end{equation}

In the opposite limit of a size-constrained grating ($L \ll L_p$), the exponential may be Taylor-expanded to yield
\begin{equation}
    Q(L) \simeq \frac{\omega_0}{2\Gamma}\,2\frac{\Gamma}{v_g}L
    = \frac{\omega_0 L}{v_g} = \frac{\omega_0}{\Gamma} \frac{L}{L_p}.
\end{equation}
Remarkably, in this regime the quality factor is primarily governed by the mode's transit time to the edge, $t=L/v_g$. This implies that, for strongly footprint-constrained devices, high-$Q$ requires increasing the interaction time within the fixed footprint by engineering a reduced group velocity (slow light).

\subsection*{Interference fringes}
In addition to setting the resonance lineshape, the finite grating length also gives rise to oscillatory fringes in the $k$-space reflectance. Specifically, the region over which the guided mode can couple out to the far field is defined by the effective length $L = W - x_0$, and the resulting fringes arise from coherent interference between contributions emitted from different positions along the grating. This corresponds to multiplication of the real-space field by a rectangular window, whose Fourier transform produces a sinc-like modulation in $k$-space. As a result, the reflected intensity exhibits an interference pattern with characteristic fringe spacing
\begin{equation}
\Delta k_\mathrm{fringe} = \frac{2\pi}{L}.
\end{equation}
Owing to redistribution of spectral weight in $k$-space by this finite-aperture interference, the reflectance can locally exceed unity without implying optical gain; the scattered power remains constrained by the input power minus the dissipated power.